\documentstyle[aps,prb,preprint,graphicx,amsbsy,amssymb]{revtex}
\tightenlines
\begin{document}


\title{Imaging the structure of the domain wall between symmetries
interconnected
by a discontinuous transition}

\author{Yanina Fasano, M. De Seta$^{*}$, M. Menghini, H. Pastoriza and F. de la
Cruz}
\address{Instituto Balseiro and Centro At\'omico Bariloche, CNEA, Av. Bustillo
9500, Bariloche, RN, Argentina}

\date{\today}
\maketitle

\begin{abstract}

We have been able to observe with single particle resolution the interface
between two structural
symmetries that cannot be interconnected by a continuous transition. By means
of an engineered 2D
potential that pins the extremity of vortex strings an square symmetry was
imposed at the surface of a
3D vortex solid. Using the Bitter decoration technique and on account of the
continuous vortex symmetry
we visualize how the induced structure transforms along the vortex direction
before changing into the
expected hexagonal structure at a finite distance from the surface.

\end{abstract}



First order phase transitions\cite{1,2} are often associated with a
discontinuous change of topological symmetry.
The analysis and detection at microscopic scale of mechanisms that allow the
nucleation and propagation of new symmetries is a formidable task of difficult
experimental resolution.
An ensemble of superconducting vortices, having a continuous symmetry along the
average vortex orientation and discrete structural symmetry in planes
perpendicular to it, is a unique toy model to investigate first and second
order phase transitions\cite{3,4,5,6,7,8,9}.

The vortex liquid in superconducting $Bi_{2}Sr_{2}CaCu_{2}O_{8}$ crystals
transforms through a first order thermodynamic transition into a solid ( Bragg
Glass\cite{12} ) with hexagonal quasi-long range order.
As a result of the almost ideal behavior of the vortex structure  large vortex
single crystals are usually obtained after the liquid-solid transition.
The topology of the solid state with individual vortex resolution can be
detected by magnetic decoration, observing the  clumps generated by small Fe
particles deposited at the vortex positions.

The scope of this work is to develop a strategy to investigate how a change of
structural symmetry induced at the surface of a solid of elastic strings
propagates along its length. This is achieved imposing the vortex liquid to
solidify in a three dimensional square lattice at one surface of a
$Bi_{2}Sr_{2}CaCu_{2}O_{8}$ crystal while the natural hexagonal vortex
structure is observed at the opposite surface.
In this way, we were able to study the vortex structure within the interface
between two single crystals of vortices with symmetries that cannot be
interconnected by a continuous phase transition.

By means of electron beam lithography Fe dots of  300\,nm in diameter and 60\,
nm height were deposited on the top surface of $Bi_{2}Sr_{2}CaCu_{2}O_{8}$
single crystals of approximate dimensions 0.5\,x\,0.5\,x\,0.03\,mm$^3$.
The dots are of similar dimensions as the clumps generated in magnetic
decorations that act as effective surface pinning centers (Bitter pinning\cite{
13}).
In this way periodic Fe patterns of square symmetry with 50\,x\,50  dots were
created. Large surface regions of the sample were left free of Fe dots to
detect the hexagonal pattern and to compare the induction field $B$ inside and
out of the Fe patterned region. The vortex structure was nucleated by cooling
the sample in the presence of a magnetic field. The Fe pattern was chosen to
make the area of its unit cell,  $a_{sq}^2$, equal to that of the corresponding
hexagonal vortex lattice induced by the applied field ( one flux quantum per
unit cell ).
The vortex lattice was visualized by magnetic decoration.
The Fe structure from the magnetic decoration, as well as that of the periodic
pattern, was observed at room temperature with a scanning electron microscope.
When the Bitter and the periodic pattern coincide, the Fe clumps from the
decoration were recognized by their irregular appearance as compared to the
circular shape of the dots of the periodic pattern.
The number of vortices per unit area in the patterned area was found to be the
same as that in the non-patterned region ( the same $B$).

\begin{figure}[htb]
\includegraphics[width=85mm]{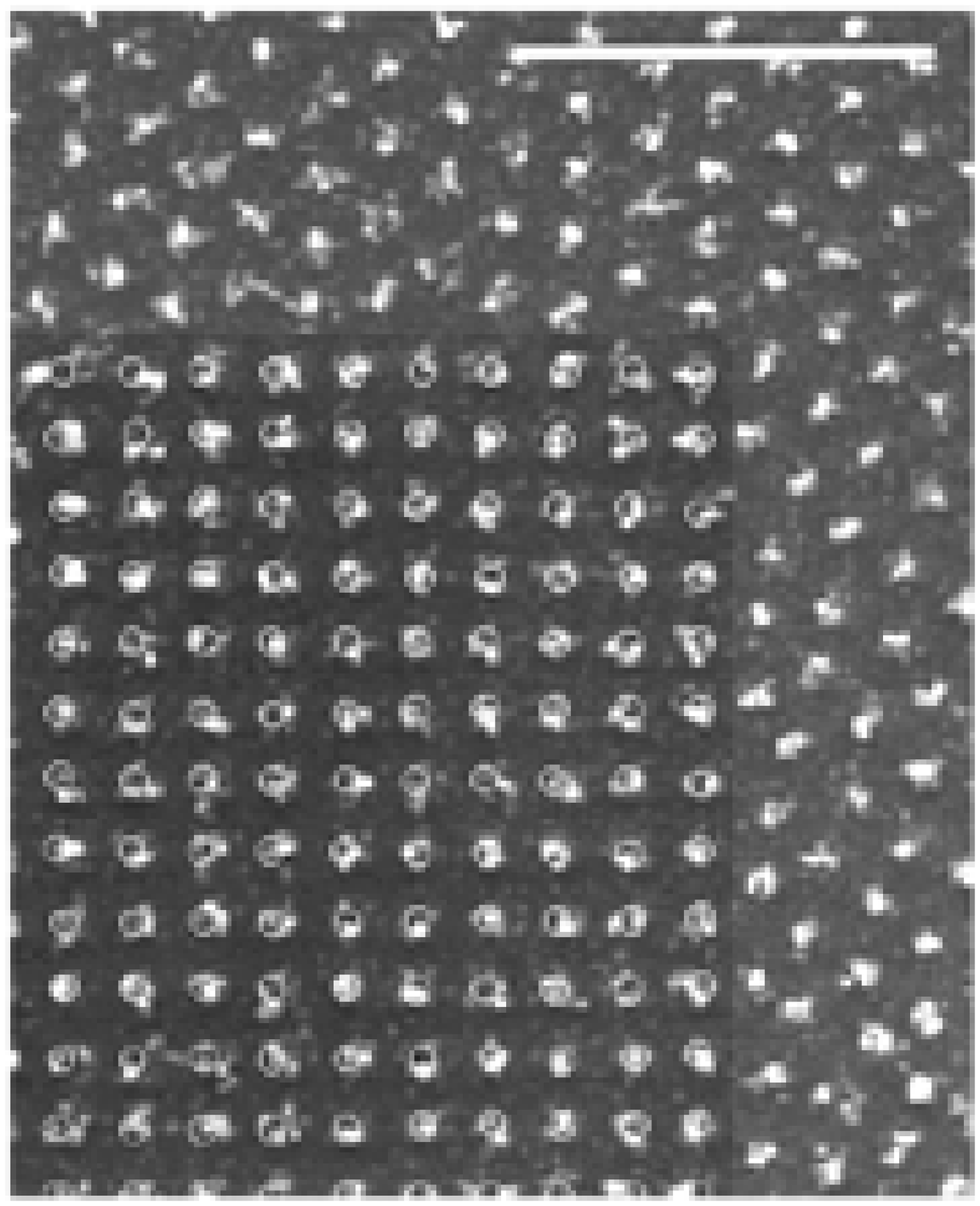}
\caption[]{Vortex decoration of the top surface of a sample with commensurate
square magnetic pattern for a field of 30.2\,Gauss at 4.1\,K.  The Fe dots of
the periodic structure are 260\,nm in diameter and 60\,nm in height. The dark
area indicates the region where a periodic $2D$ square pattern was generated.
The Fe dots positions are depicted with white circles. The scale bar
corresponds to 5\,$\mu$m.}
\label{fig:1}
\end{figure}

Figure\ \ref{fig:1} shows the vortex decoration on the Fe-patterned surface of
a sample with $a_{sq} = 0.835$\,$\mu$m, hereafter called top surface.
The vortex structure corresponds to  $B = 30.2$\,Gauss and the decoration was
made at 4.1\,K.
The dark region marks the area where the square $2D$ pinning was generated.
The square topology of the vortex structure induced by the periodic pattern is
evident. The natural hexagonal symmetry of the solid is seen to be recovered
outside the Fe patterned region in no more than two lattice parameters.
The one to one site correspondence of Fe clumps from the decoration with the
square pattern reveals that the $2D$ periodic pinning imposes the vortex liquid
to solidify in a single crystal with square symmetry at the top surface.

\begin{figure}[htt] 
\includegraphics[width=85mm]{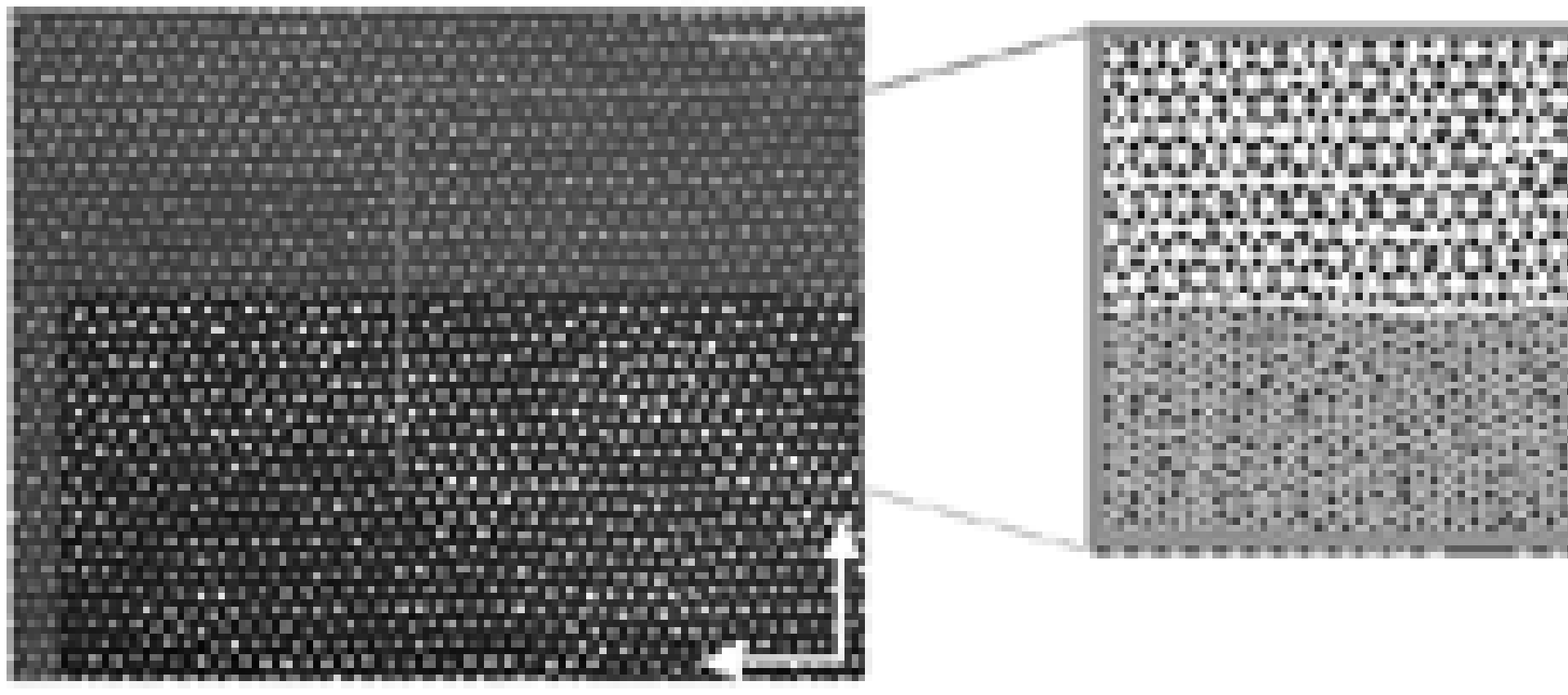}
\caption[]
{Vortex decoration of the bottom surface of a 4.5\,$\mu$m thick sample beneath
the square Fe pattern
indicated by the dark region with brighter spots. The insert depicts the
corresponding Delaunay triangulation of the  framed
region, obtained joining the positions of next-near-neighbors vortices with
straight lines. The picture
shows that the hexagonal vortex lattice has one of its principal directions
parallel to one of the
square pinning potential axis (indicated by the arrows). The scale bar
corresponds to 5\,$\mu$m.}
\label{fig:2} \end{figure}

To verify whether the sample is thick enough to allow the recovery of the
hexagonal vortex symmetry, a
magnetic decoration was made on the opposite surface of the Fe patterned one,
hereafter called bottom
surface. Figure\ \ref{fig:2} shows the decoration at the bottom surface of a
4.5\,$\mu$m thick sample,
beneath the square Fe patterned region. The Delaunay triangulation of the
vortex lattice is depicted.
The picture shows that the hexagonal vortex lattice has one of its principal
directions parallel to one
of the square pinning potential axis. This shows that the presence of the
square potential at the top
surface breaks the rotational degeneracy of the hexagonal lattice. The hexagons
at the bottom of the
sample make evident the coexistence of square and hexagonal symmetries along
the vortex strings.

The results in Fig.\ \ref{fig:1} and\ \ref{fig:2} suggest the presence of a
domain wall between structures with equivalent energy: The square structure at
one side of the wall, where the surface pinning potential compensates the
increase of interaction energy corresponding to the square lattice, and the
natural hexagonal symmetry of the vortex crystal at the other side.

To explore the change of the vortex topology along the sample thickness,
decorations of freshly cleaved bottom surfaces of the sample were performed.
The cleaving method has the advantage of preparing excellent surfaces for
decoration but does not allow us to choose a precise thickness.
Repeating the cleaving process we were able to change the thickness of the
original sample of 12.5\,$\mu$m.
Decoration of the bottom surface of samples with thickness, d, of 12.5, 5.75,
4.5, 3.5, 2.5 and 0.5\,$\mu$m  verified that the hexagonal symmetry was
recovered at a distance between 3.5\,$\mu$m and 4.5\,$\mu$m from the top
surface.

\begin{figure}[htt]
 \includegraphics[width=88mm]{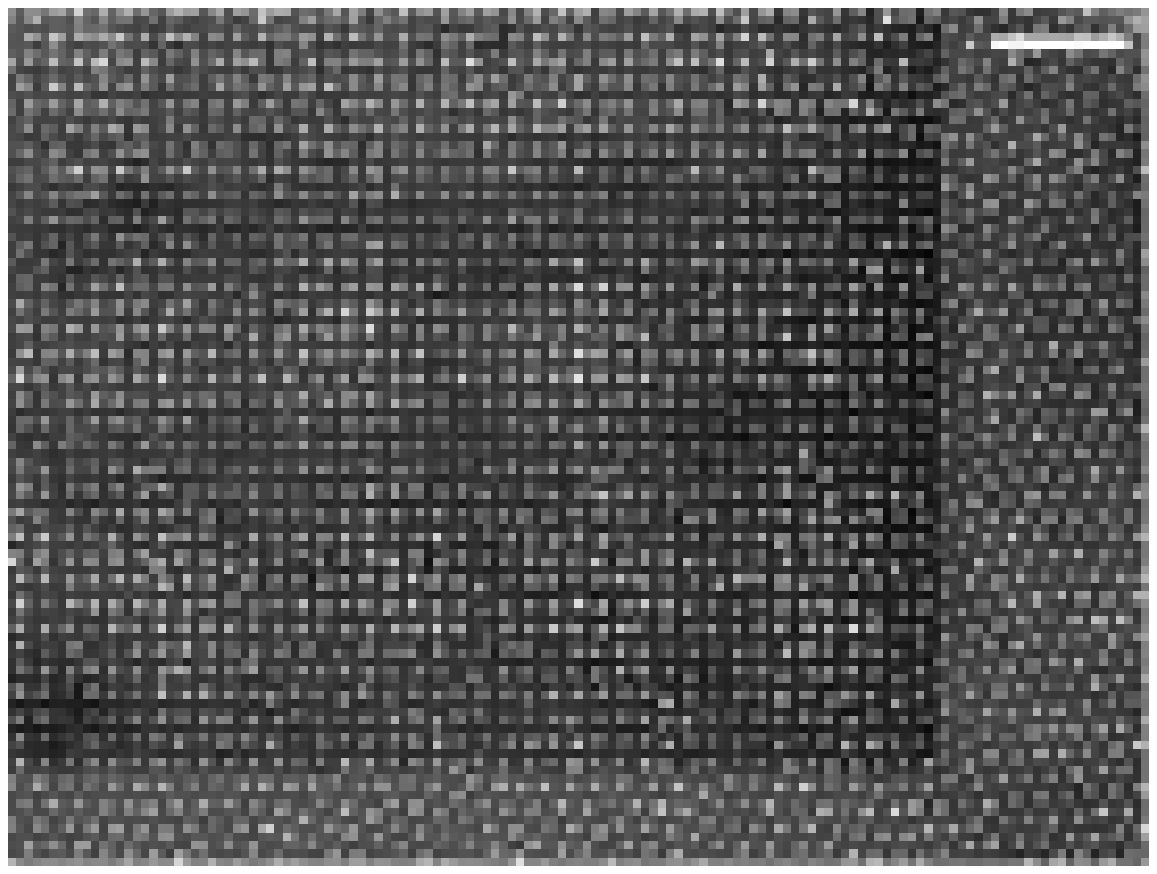}
\caption[]{Vortex decoration of the bottom surface of the thinnest sample
 investigated, $d = 0.5$\,$\mu$m. It shows a complete square symmetry of the
 whole vortex structure beneath the periodic $2D$ square pattern in the top
 surface ( dark area ). The scale bar corresponds to 5\,$\mu$m. }
 \label{fig:3}
\end{figure}

The thinnest sample, with $d = 0.5$\,$\mu$m, shows that the vortex structure
has the square symmetry in the whole region beneath the Fe pattern, as depicted
in Fig.\ \ref{fig:3}.
This result indicates that the engineered $2D$ pinning potential can be used to
modify the melting or wetting conditions for the vortex system in a finite
thickness close to the surface of the sample.

The magnetic decoration of vortices at the bottom surface of samples with
different thicknesses allowed to locate the domain wall between the square and
hexagonal symmetries and to investigate its internal microscopic structure.
Since  the interface is detected close to the top surface, starting somewhere
between 0.5\,$\mu$m and 2.5\,$\mu$m from the top surface and extends to a
distance between 3.5\,$\mu$m and 4.5\,$\mu$m from it, the domain wall has a
thickness of the order of the average vortex near neighbors distance (a
$\approx$ 1\,$\mu$m).

\begin{figure}[htt]
\includegraphics[]{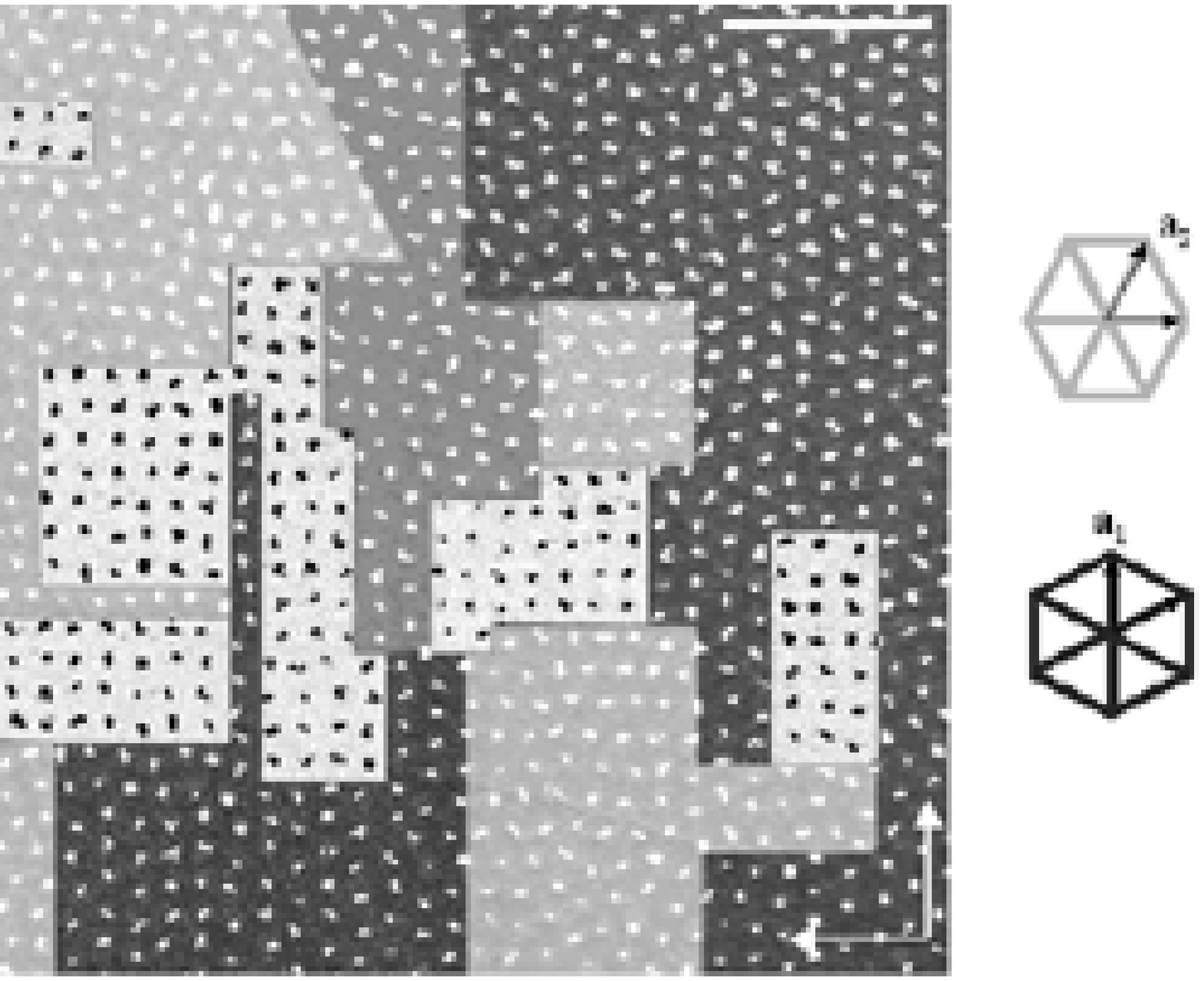}
\caption[]{Image of the vortex topology in the interface. Decoration of the
bottom surface of the 3.5\,$\mu$m thick sample. Arrows indicate the principal
directions of the square pattern on the top surface. Three distorted hexagonal
symmetries coexisting with fourfold symmetry regions are shown. Black clumps on
white: Domains of vortices in square symmetry matching the square Fe pattern at
the top surface. White clumps on light and dark gray: Vortices in distorted
hexagons with unit cell area equal to the square pattern unit area (
$a_{1} = a_{sq}$ parallel to one of the principal directions of the pattern and
$a_{2} =1.11a_{sq}$ ). White clumps on gray: Vortices in distorted hexagons
with unit cell area equal to the square pattern unit area, rotated 15\,degrees
with respect to the horizontal principal direction of the square pattern. The
scale bar corresponds to 5\,$\mu$m.
}
\label{fig:4}
\end{figure}

Decoration of vortex structures at the bottom of  2.5\,$\mu$m  and 3.5\,$\mu$m
thick samples have shown a distribution of  three distorted hexagonal
structures coexisting with regions containing a small number of vortices ( 15
in average ) with fourfold symmetry which lattice parameter is equal to that of
the square Fe pattern on the top surface.
As an example, Fig.\ \ref{fig:4} shows the decoration of one pattern obtained
at the bottom of 3.5\,$\mu$m sample.
Statistics made on hundreds of vortex unit cells below the square Fe pattern
indicate that from the total number of vortices in the domain wall 83\% belong
to hexagonal deformed structures, while the rest are distributed in regions of
square symmetry of few lattice parameters.

\begin{figure}[htt]
\includegraphics[]{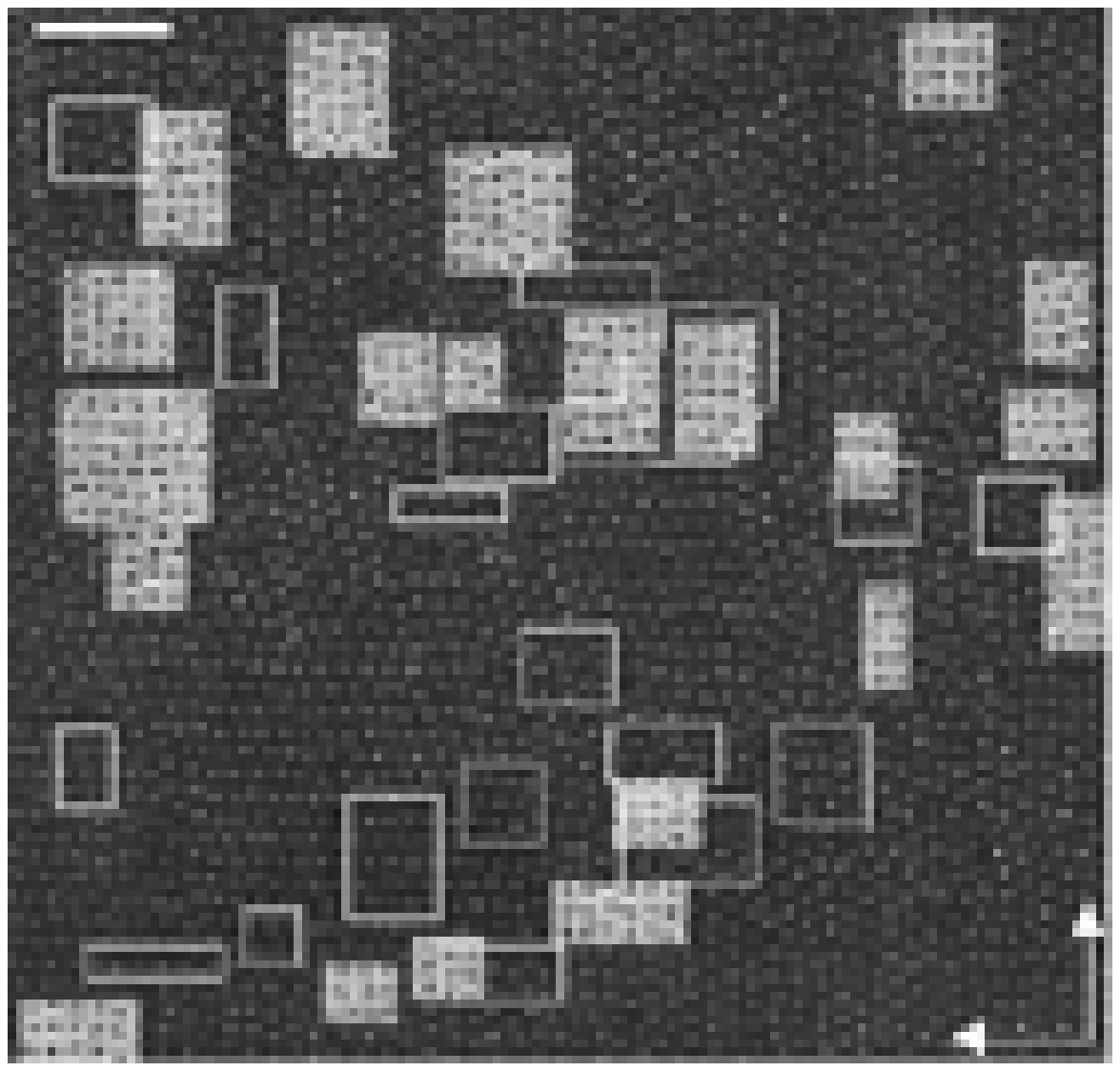}
\caption[]{Comparison of vortex interface structures at 3.5 and 2.5\,$\mu$m
underneath the same square Fe pattern. The picture is that of the 3.5\,$\mu$m
sample where the black clumps  areas depict the square symmetry regions and the
superimposed white frames indicate the location of the square symmetry regions
in the 2.5\,$\mu$m thick sample. The scale bar corresponds to 5\,$\mu$m.}
\label{fig:5}
\end{figure}

The structure at the bottom of 2.5\,$\mu$m thick sample shows the same relative
distribution (within 3\%) of distorted hexagonal and square symmetry regions.
It is clear that at these two different thicknesses (one close to the beginning
and the other to the end of the domain wall) the interface embraces the
coexistence of deformed hexagons and square vortex structures.
Similar configurations were found by recent numerical simulations of vortex
interfaces between square and hexagonal symmetries \cite{fabiana}.
A fraction of 85\% of the deformed hexagons are distributed between two
possible degenerate states rotated in 90\,degrees.
Each state is characterized by one lattice parameter equal in magnitude and
parallel to one of the square Fe pattern, see Fig.\ \ref{fig:4}.
The results from four decorations made at the bottom of 3.5\,$\mu$m and 2.5\,
$\mu$m  thick samples indicate that, despite the proportion of square and
deformed hexagons is maintained, the relative abundance between the two
hexagonal degenerate states differs widely from one experiment to another.
The comparison among vortex structures at 3.5 and 2.5\,$\mu$m underneath the
same square Fe pattern shows that the specific location of square and distorted
hexagonal vortex regions varies from one experiment to the other, see Fig.\
\ref{fig:5}.
This strongly supports that the proportion of deformed hexagons and square
symmetries is an intrinsic characteristic of the interface between the square
and hexagonal single crystals nucleated along the vortex string.

Another characteristic of the interface is that the area of both, the hexagonal
distorted as well as the square unit cell at the bottom of the 3.5 and 2.5\,
$\mu$m samples coincides within 2\% with that of  the square and hexagonal
crystals at each side of the domain wall.
This strongly supports that the transformation from the square to the hexagonal
symmetry is made under the condition of constant unit cell area (uniform $B$)
across the interface. The uniaxial compression of the hexagons results from a
distortion induced to match one of the hexagonal lattice vector with one of the
square pattern on the top surface.
The transformation from the described mixture of phases within the interface
region to the uniform hexagonal symmetry takes place within one micrometer (one
lattice parameter).
We have not observed this transformation because the cleaving technique does
not permit to choose a thickness of the sample with the necessary precision .

In conclusion, the combination of electron-beam lithography and magnetic
decoration techniques have allowed to demonstrate that the vortex liquid can be
solidified in two coexisting crystals with different symmetries along the
vortex length.
The square symmetry induced by the engineered $2D$ Fe pattern at one end of the
vortex strings propagates up to a distance of the order of a micrometer and
converts into a hexagonal lattice through a domain wall.
The square symmetry is seen to transform across the  domain wall creating large
regions of distorted hexagonal symmetry coexisting with others of square
symmetry of few lattice parameters size.
The continuous nature of the vortex strings is the unique property that allows
us to visualize the transformation of the structure along the vortex direction
in a transition width of the order of the average separation between vortices.
The experimental data provide valuable information to stimulate and verify the
results of theoretical modeling of interfaces.

\section{Acknowledgments} We acknowledge A. A. Aligia, J. Lorenzana, C.
Balseiro, E. Jagla, P. Cornaglia
and F. Laguna for valuable discussions and E. Martinez for careful reading of
the manuscript. This work
was partially supported by Agencia Nacional de Promoción Científica y
Tecnológica, by Fundación
Antorchas and by Consejo Nacional de Investigaciones Cientificas y Técnicas,
Argentina.

\begin{references}

\bibitem[]{byline} {*} Permanent address: Unità INFM e Dipartimento di Fisica
Universitá di Roma Tre,
Rome, Italy.


\bibitem{1} Landau, L. D. and Lifshitz, E. M., {\it Statistical Physics}  (
Pergamon, Oxford, 1993)3rd edn.

\bibitem{2} Chaikin, P. M. and Lubensky, T. C., {\it Principles of condensed
matter physics} (Cambridge University Press, Cambridge, 1995)1st edn .

\bibitem{3} H. Safar, P. L. Gammel, D. A. Huse, D. J. Bishop, J. P. Rice
and  D. M. Ginsberg, Phys. Rev. Lett. {\bf 69}, 824 (1992).

\bibitem{4} Pastoriza, H., Goffman, M. F., Arribére, A. and de la Cruz, F.,
Phys. Rev. Lett. 72, 2951 (1994).

\bibitem{5} Zeldov, E. et al., Nature 375, 373 (1995).

\bibitem{6} Nelson, D. R. and Vinokur, V. M., Phys. Rev. B  48, 13060 (1993).

\bibitem{7} Grigera, S. A. , E. Morre, E. Osquiguil, C. Balseiro, G. Nieva, and
F. de la Cruz,
Phys. Rev. Lett. 81, 2348 (1998).

\bibitem{8} Blatter, G. et al., Rev. Mod. Phys. 66, 1125 (1994).

\bibitem{9} Fisher, D. S., Fisher, M. P. A. and Huse, D. H., Phys. Rev. B  43
130 (1991).

\bibitem{12} T. Giamarchi and P. Le Doussal, Phys. Rev. B
{\bf 52}, 1242 (1995).

\bibitem{13} Yanina Fasano, J. A. Herbsommer, F. de la Cruz, F. Pardo, P. L.
Gammel, E.
Bucher and D. J. Bishop, Phys. Rev. B {\bf 60}, R15047 (1999).

\bibitem{fabiana}  P. Cornaglia and M. F. Laguna (private communication).


\end{references}
\end{document}